\documentclass[5p]{elsarticle}
\usepackage{graphicx}
\usepackage{epsfig,latexsym}

\newcommand{\be}{\begin{equation}}
\newcommand{\ee}{\end{equation}}
\newcommand{\ber}{\begin{eqnarray}}
\newcommand{\eer}{\end{eqnarray}}

\begin{document}
\title{Beam energy dependence of Elliptic and Triangular flow with the AMPT model}

\author[ens]{Dronika Solanki}
\ead{dronika@rcf.rhic.bnl.gov}
\author[bnl]{Paul Sorensen}
\ead{psoren@bnl.gov}
\author[pri]{Sumit Basu} 
\ead{sumit.basu@cern.ch}
\author[ens]{Rashmi Raniwala} 
\author[pri]{Tapan Kumar Nayak}
\cortext[cor1]{Corresponding author}

\address[ens]{Physics Department, University of Rajasthan, Jaipur 302004, India}
\address[bnl]{Brookhaven National Laboratory, Physics Department, Upton, NY 11973, USA}
\address[pri]{Variable Energy Cyclotron Centre, Kolkata 700064, India}

\begin{abstract} A beam energy scan has been carried out at the
  Relativistic Heavy Ion Collider at Brookhaven National Laboratory to
  search for the onset of deconfinement and a possible critical point
  where the transition from a Quark Gluon Plasma to a hadronic phase
  changes from a rapid cross-over to a first order phase
  transition. Anisotropy in the azimuthal distribution of produced
  particles such as the second and third harmonics $v_2$ and $v_3$ are
  expected to be sensitive to the existence of a Quark Gluon Plasma
  phase and the Equation of State of the system. For this reason, they
  are of great experimental interests.  In this Letter we report on
  calculations of $v_2$ and $v_3$ from the AMPT model in the Default(Def.)
  and String Melting(SM) mode to provide a reference for the energy
  dependence of $v_2$ and $v_3$ for $\sqrt{s_{_{NN}}}$ from 7.7 GeV to
  2.76 TeV. We expect that in the case that collisions cease to
  produce QGP at lower colliding energies, data will deviate from the
  AMPT String Melting calculations and come in better agreement with the
  Default calculations.  \end{abstract}


\maketitle

\section{Introduction}

The motivation for colliding heavy ions at facilities like the
Relativistic Heavy Ion Collider (RHIC) at Brookhaven National
Laboratory and the Large Hadron Collider (LHC) at CERN is to form a
state of matter called the Quark Gluon Plasma
(QGP)~\cite{Reisdorf:1997fx}. Each of these collisions creates a
region so hot and dense that quarks and gluons become the relevant
degrees of freedom instead of hadrons~\cite{eos}. Studying the
conversion of coordinate space anisotropies into momentum space
anisotropies gives insight into the nature of the matter created in
these collisions~\cite{Voloshin:2008dg}. For decades, elliptic flow
($v_2=\langle\cos2(\phi-\Psi_{\mathrm{RP}})\rangle$) has been studied to probe
the conversion of the elliptic shape of the initial overlap zone into
azimuthal anisotropy in 
momentum space~\cite{v2papers} over a broad range of colliding beam
energies. Measuring the strength of that conversion as a function of
beam energy to search for evidence of the onset of deconfinement or a
softening of the equation-of-state is one of the goals of the RHIC
Beam Energy Scan program. In 2007 Mishra et.  al.~\cite{Mishra:2007tw}
proposed the analysis of $v_n^2$ for all values of $n$ and argued that
density inhomogeneities in the initial state would lead to non-zero
$v_n^2$ values for higher harmonics including $v_3$. Although they
proposed that $v_n$ vs. $n$ could be used to search for superhorizon
fluctuations, it was later noted that higher harmonics of $v_n$ would
be washed out by viscous effects and that the shape of $v_n$ vs. $n$
would provide a valuable tool for studying $\eta/$s~\cite{sound}. It
was also subsequently pointed out that information on $v_n^2$ was to a
large extent contained within already existing two-particle
correlations data~\cite{Sorensen:2008dm}, and that $v_n$ and $v_n$
fluctuations would provide a natural explanation for the novel
features seen in those correlations, such as the ridge
like~\cite{ridgedata} and mach-cone like~\cite{Adams:2005ph}
structures. That the ridge could be related to flux-tube like
structures in the initial state was already argued by Voloshin in
2006~\cite{Voloshin:2004th}. Calculations carried out within the
NEXSPHERIO model showed that in a hydrodynamic model fluctuations in
the initial conditions lead to a near-side ridge correlation and a
mach-cone like structure on the away-side~\cite{Takahashi:2009na}. In
2010, Alver and Roland used a generalization of participant
eccentricity ($\varepsilon_{n,\mathrm{part}}$) to arbitrary values of
$n$ as in Ref.~\cite{Broniowski:2007ft} and showed that within the
AMPT model, the final momentum space anisotropy for $v_3$ was
proportional to the initial
$\varepsilon_{3,\mathrm{part}}$~\cite{Alver:2010gr}. This explained
the previous observation that the AMPT model produced correlations
similar to those seen in the data (albeit with smaller
amplitudes)~\cite{Ma:2006fm}. Later studies showed that with changes
to the input parameters, AMPT could quantitatively describe the
centrality dependence of $v_2$ and $v_3$ at 200 GeV and 2.76 TeV~\cite{Xu:2011fe}.

In this paper we use the AMPT model to study the beam energy
dependence of $v_2$ and $v_3$. We study collisions ranging from
$\sqrt{s_{_{NN}}}=$ 7.7 GeV to 2.76 TeV. We compare AMPT in the string
melting and the default setting.

\section{AMPT Settings and Comparisons to Data}

\begin{figure}[htb]
\centering
\resizebox{0.49\textwidth}{!}{\includegraphics{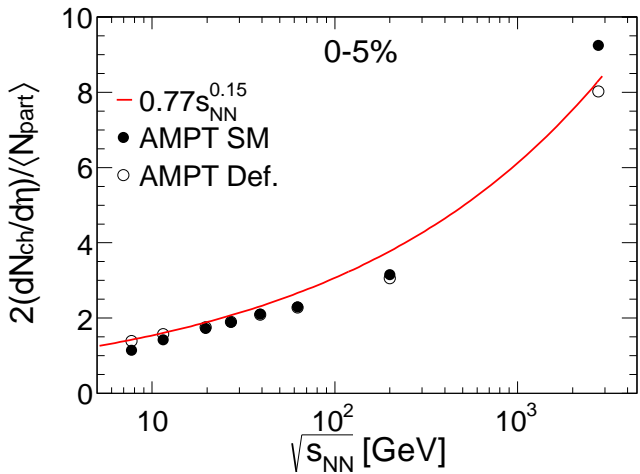}}
\caption[]{ The charged particle multiplicity density scaled by $N_{\mathrm{part}}/2$ in the AMPT model for String Melting  and Default modes. The red line shows the parameterization of experimental data presented in Ref~\cite{ALICE_mult}. }
\label{f1}
\end{figure}

The AMPT model provides two modes: Default and String Melting
\cite{Lin:2004en}. AMPT in default mode is essentially a string and
minijets model (without a QGP phase) where initial strings and
minijets are produced with the HIJING event
generator~\cite{Wang:1991hta}. The interactions of the minijet partons
are then calculated using a parton cascade (ZPC) before the strings
and partons are recombined and the strings are fragmented via Lund
string fragmentation. ART (A Relativistic Transport model for
hadrons) is used to describe how the produced hadrons will
interact. In the String Melting mode, the strings produced from HIJING
are decomposed into partons which are fed into the parton cascade
along with the minijet partons. The partonic matter is then turned
into hadrons through coalescence and the hadronic interactions are
subsequently modeled using ART. So while the Default mode describes
the collision evolution in terms of strings and minijets followed by
string fragmentation, the String Melting mode includes a fully
partonic QGP phase that hadronizes through quark coalescence. The
model therefore provides a convenient way to investigate expectations
for a variety of observables with and without a QGP phase.

Several parameters need to be specified in the model including
parameters a and b for Lund string fragmentation, the QCD coupling
constant $\alpha_s$ (which the model treats as a constant), and the
screening mass for gluons in the QGP phase $\mu$. A recent study found
that a good description of the multiplicity density, $v_2$ and $v_3$
could be achieved with the parameter set: a=0.5, b=0.9 (GeV$^{-2}$),
$\alpha_s$=0.33 and $\mu$=3.2 (fm$^{-1}$)~\cite{Xu:2011fe}. In this
study, we found that we can acheive a good desciption of the
multiplicity density at all energies from $\sqrt{s_{_{NN}}}$= 7.7 GeV to
2.76 TeV by using parameter set: a=2.2, b=0.5 (GeV$^{-2}$),
$\alpha_s$=0.47 and $\mu$=1.8 (fm$^{-1}$) and turning off initial and
final state radiation in HIJING. In this case, the initial cutoff for
minijets $p_0$ does not need to be adjusted with $\sqrt{s}$ in order to
match the LHC multiplicity densities~\cite{Deng:2010mv}. We leave
$p_0$ and all other parameters fixed for all energies. Figure~\ref{f1}
shows the charged particle multiplicity density scaled by $N_{\mathrm{part}}/2$
for 0-5\% central Au+Au or Pb+Pb collisions from AMPT String Melting
and Default vs $\sqrt{s_{_{NN}}}$. The line shows the
parameterization of the experimental data from
Ref.~\cite{ALICE_mult}. Both the SM and Default calculations are in
good agreement with the experimental data throughout the entire energy
range.

\begin{figure}[htb]
\centering
 \resizebox{0.49\textwidth}{!}{\includegraphics{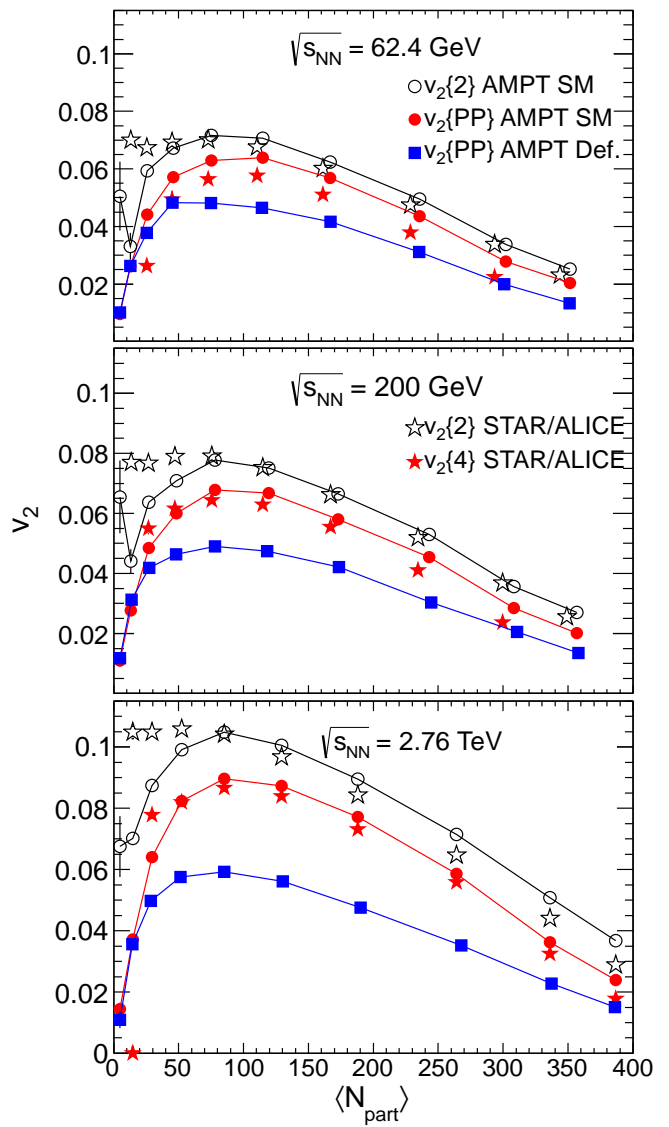}}
 \caption[]{ Elliptic flow data from AMPT and experiments at
   $\sqrt{s_{_{NN}}}$=62.4 GeV, 200 GeV [STAR],  and 2.76 TeV [ALICE]. For the String
   Melting calculation we show $v_2$ calculated relative to the
   participant plane $v_{2}\{\mathrm{PP}\}$ defined by the positions of the
   nucleons and using the two particle cumulant
   $v_2\{2\}=\langle\cos2(\phi_i-\phi_j)\rangle$. Experimental results
   are shown for the two-particle $v_2\{2\}$ and four-particle
   $v_{2}\{4\}$ cumulants.}
\label{f2}
\end{figure}

In Figure~\ref{f2}, the AMPT model results are compared to
experimental data at $\sqrt{s_{_{NN}}}=$62.4 GeV, 200 GeV, and 2.76
TeV. For the SM calculations, we show (i) $v_2$ relative to the
participant plane ($v_2\{PP\}$) calculated from the initial conditions
of AMPT and (ii) the two-particle cumulant results $v_{2}\{2\}$.
While $v_{2}\{2\}=\sqrt{\langle v_2^2\rangle} + \delta$ where $\delta$
is a term to account for correlations not related to the participant
plane (non-flow), $v_{2}\{P.P.\}$ is the true mean $v_2$ relative to
the participant plane.  The difference between those results therefore
reflects both the effect of fluctuations
$\sqrt{\langle v_2^2\rangle}-\langle v_2\rangle^2$ and non-flow
correlations present in the model.  All the model calculations have a
similar centrality dependence but the Default results are well below
the SM results. The data generally agree well with the SM
calculations. In the case that the $v_2$ fluctuations are dominated by
eccentricity fluctuations and those eccentricity fluctuations follow a
Gaussian distribution in $x$ and $y$, $v_2\{4\}$ should be equal to
$v_2$ with respect to the reaction plane~\cite{gmod}. The fact that
the experimental $v_2\{4\}$ results are slightly below the model
results calculated with respect to the participant plane does not
therefore signify a discrepancy between data and model. We consider
the agreement between the model and the data to be satisfactory.

\begin{figure}[htb]
\centering
\resizebox{0.5\textwidth}{!}{\includegraphics{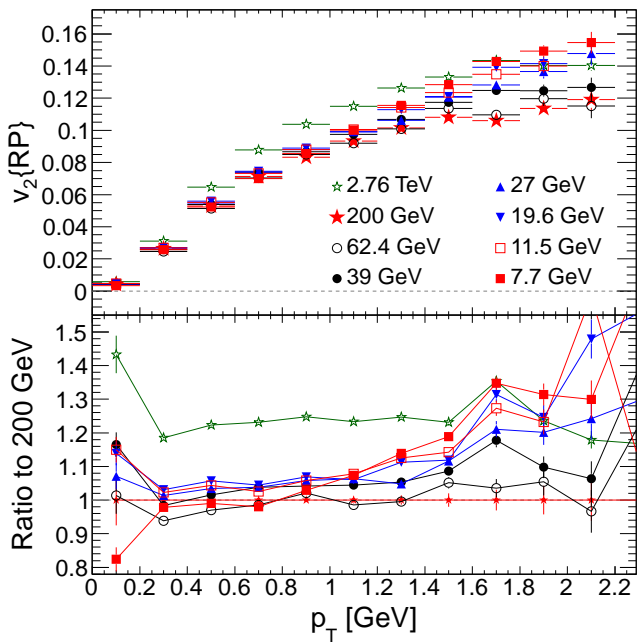}}
\caption[]{ Top: $v_2(p_T)$ calculated with AMPT SM relative to the
  reaction plane ($v_2\{\mathrm{RP}\}$) for beam energies from 7.7 GeV to 2.76
  TeV. Bottom: The $v_2\{\mathrm{RP}\}$ data at different energies are shown
  scaled by the results at 200 GeV.}
\label{f7}
\end{figure}

STAR has shown that for $p_T<1$ GeV, $v_2\{4\}(p_T)$
increases with $\sqrt{s_{_{NN}}}$, for $p_T>1$ GeV $v_2\{4\}(p_T)$ is
roughly independent of collision energy in the range 7.7 GeV to 2.76
TeV~\cite{STARBESv2}. It is surprising for a measurement that is
supposed to be sensitive to viscosity and collective effects in the
expansion to not depend on $\sqrt{s_{_{NN}}}$ over such a wide range of
energies where the initial conditions and properties of the fireball
should be changing quite significantly. Given this surprising
experimental result, it is interesting to see if the same trend is
reproduced in the AMPT model. In Figure~\ref{f7}, we show $v_2(p_T)$
calculated with respect to the reaction plane for collisions with center
of mass energies ranging from 7.7 GeV to 2.76 TeV. Although the
statistics in our study were not sufficient to calculate
$v_2\{4\}(p_T)$, it has been shown that as long as $v_2$ fluctuations
are dominated by eccentricity fluctuations and those eccentricity
fluctuations are Gaussian distributed along the x and y axis, then
$v_2\{4\} $ is equivalent to $v_2\{\mathrm{RP}\}$~\cite{gmod}. We therefore
check to see if $v_2\{\mathrm{RP}\}(p_T)$ is independent of $\sqrt{s_{_{NN}}}$
for $p_T>1$ GeV in the AMPT model. We find that the variation of
$v_2\{\mathrm{RP}\}$ is not large in AMPT throughout the energy range
studied. For $p_T<1$ GeV, $v_2\{\mathrm{RP}\}$ varies by about 5\% from 7.7 GeV
up to 200 GeV. Going from 200 GeV to 2.76 TeV, $v_2\{\mathrm{RP}\}$ increases by
20\%, independent of $p_T$. In the RHIC range, the AMPT $v_2\{\mathrm{RP}\}$
results for $p_T>1$ GeV are actually increasing as the energy is
decreased with $v_2\{\mathrm{RP}\}$ at $p_T=1.5$ GeV for 7.7 GeV being 20\%
larger than for 200 GeV. This likely reflects the softening of the
spectrum which allows flow effects that push low momentum particles to
higher momentum, to have a larger influence at intermediate $p_T$. The
same trends hold when studying $v_2\{\mathrm{PP}\}(p_T)$ (not shown). Although
there are differences between the trends seen in AMPT and in the data,
one can conclude that even in the AMPT model, the changes in
$v_2\{\mathrm{RP}\}(p_T)$ or $v_2\{\mathrm{PP}\}(p_T)$ when increasing
$\sqrt{s_{_{NN}}}$ from 7.7 GeV to 2.76 TeV are not large. In this
case, it is not necessarily surprising that the data also does not
change drastically. Since based on the AMPT model, we would not expect
a large variation of $v_2\{\mathrm{RP}\}(p_T)$ with $\sqrt{s_{_{NN}}}$, as long
as one assumes that a string melting or QGP phase exists throughout
the energy range under study, the fact that the data seem to change
very little  no longer appears to be so difficult to understand.

\begin{figure}[htb]
\centering
\resizebox{0.45\textwidth}{!}{\includegraphics{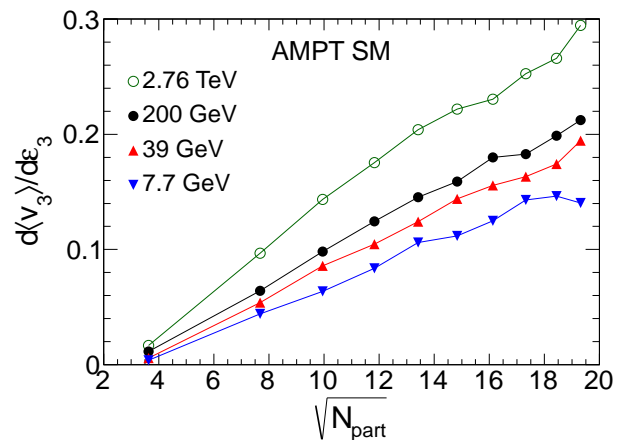}}
\caption[]{ The slope of $\langle v_3\rangle$ vs. $\varepsilon_{3}$ as
  a function of the square root of the number of participants for four
  different colliding energies.}
\label{f3}
\end{figure}

\section{The Third Harmonic}

 \begin{figure*}[htb]
\centering
 \resizebox{1.0\textwidth}{!}{\includegraphics{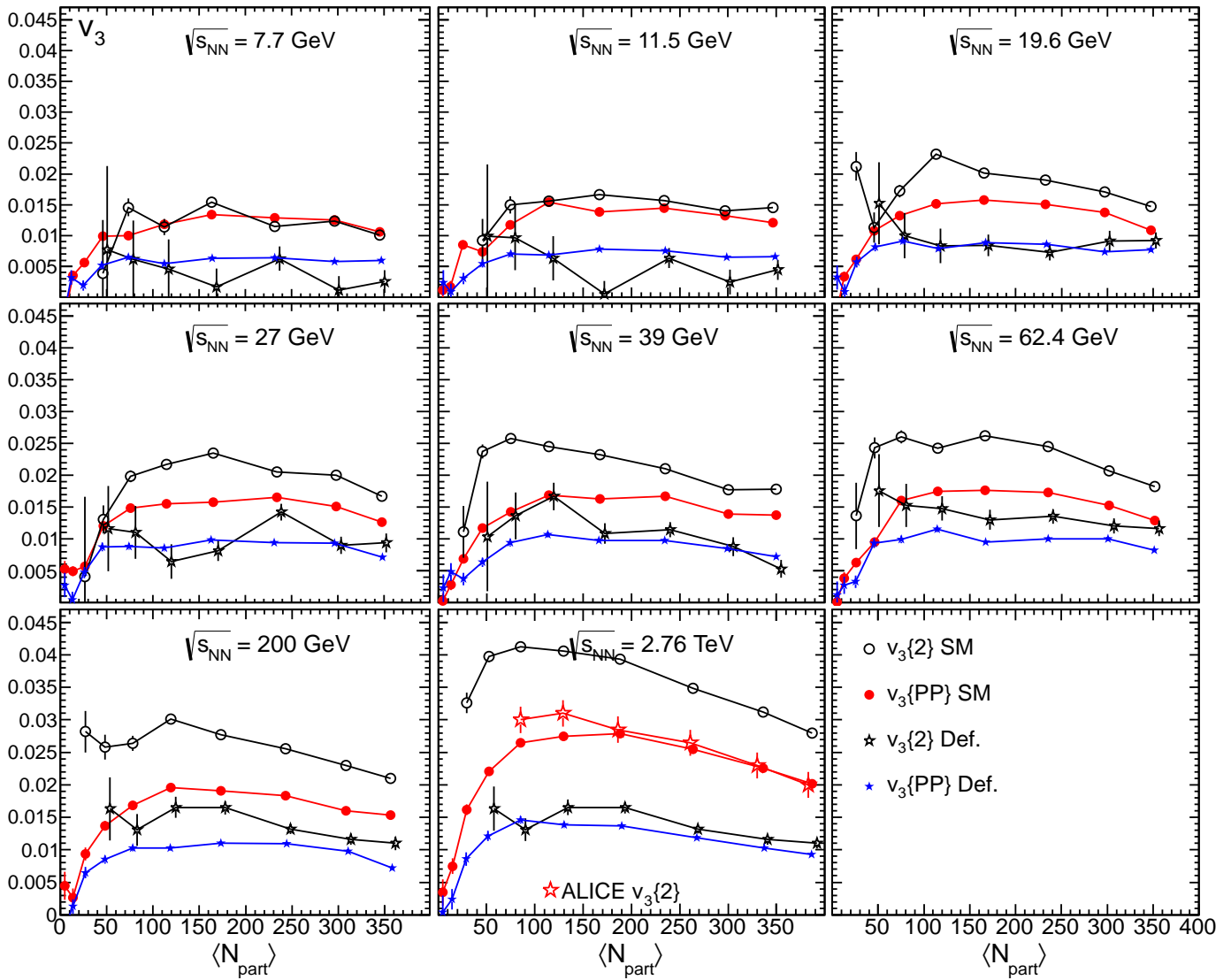}}
 \caption[]{ $v_3\{2\}$ and $v_3\{\mathrm{PP}\}$ from AMPT SM and
   Default calcualtions for $\sqrt{s_{{NN}}}=$ from 7.7 GeV to 2.76
   TeV. Experimental results are shown at 
   2.76 TeV~\cite{ALICE:2011ab}.}
\label{f4}
\end{figure*}

Having shown that our parameter selection provides a good description
of the charged particle multiplicity densities and the elliptic flow,
we now turn to investigate $v_3$ and its energy dependence. We first
study the relationship of $v_3$ to the third harmonic participant
eccentricity. In Ref.~\cite{Alver:2010gr} the AMPT model is used to
show that $v_2$ and $v_3$ have a linear dependendence on
$\varepsilon_{2}$ and $\varepsilon_{3}$. At Quark Matter 2011, STAR
showed that $v_3/\varepsilon_{3}$ scales with
1/$\sqrt{N_{\mathrm{part}}}$~\cite{Sorensen:2011fb}. Here we check to see if
this phenomenological observation is also reproduced in the AMPT
model.

In Figure~\ref{f3} we investigate the dependence of the slope of
$\langle v_3\rangle$ vs. $\varepsilon_{3}$ on $N_{\mathrm{part}}$. The figure
shows $d\langle v_3\rangle/d\varepsilon_{3}$ vs. $\sqrt{N_{\mathrm{part}}}$ for
$\sqrt{s_{_{NN}}}=7.7$ GeV, 39 GeV, 200 GeV, and 2.76 TeV. We find that
for all the energies investigated (including those not shown in the
figure), $d\langle v_3\rangle/d\varepsilon_{3}$ increases linearly
with $\sqrt{N_{\mathrm{part}}}$. The AMPT model therefore correctly describes
the phenomenological observation made by STAR. This also indicates
that according to the string melting version of AMPT, even at energies
as low as $\sqrt{s_{{NN}}}$=7.7 GeV, $v_3$ reflects the fluctuations in
the initial geometry of the collisions and that the centrality
dependence will remain similar at all energies although the magnitude
will change. At the lowest energies investigated here, the
contributions from jets and minijets should be negligible so they
will not contribute significantly to the centrality dependence of $v_3$.
The experimental observation of a similar centrality dependence for
$v_3$ at 7.7 and 200 GeV~\cite{Sorensen:2011fb}, therefore
strongly contradicts assertions that $v_3$ is dominated by
jet-like correlations~\cite{Trainor:2012jv}.

In Figure~\ref{f4}, AMPT SM and Default calculations of $v_3\{2\}$ and
$v_3\{\mathrm{PP}\}$ are shown for 7.7, 11.5, 19.6, 27, 39, 62.4, 200
GeV and 2.76 TeV. While $v_3\{\mathrm{PP}\}$ reflects the true
correlation of particles with the initial participant plane,
$v_3\{2\}$ includes non-flow and fluctuation effects. The difference between
$v_3\{2\}$ and $v_3\{\mathrm{PP}\}$ is large at 200 and 39 GeV while
at 7.7 GeV $v_3\{2\}$ and $v_3\{\mathrm{PP}\}$ are equivalent. This
indicates that indeed, according to AMPT SM, non-flow does not make an
appreciable contribution to $v_3\{2\}$ at 7.7 GeV. We compare the
model results to ALICE data at 2.76 TeV and find that
$v_3\{\mathrm{PP}\}$ for AMPT SM matches the ALICE data on
$v_3\{2\}$. The $v_3\{2\}$ AMPT SM results over predict the ALICE data
and the $v_3\{\mathrm{PP}\}$ AMPT Default results underpredict the
ALICE data. The $v_3\{2\}$ Default results also underpredict the ALICE
data for $N_{\mathrm{part}}>100$. The correspondence of
$v_3\{\mathrm{PP}\}$ from AMPT SM with $v_3\{2\}$ from ALICE data
means that either non-flow and fluctuations are overpredicted in AMPT or $v_3$ is
underpredicted. The 200 GeV data is also in good agreement with
preliminary STAR data~\cite{Sorensen:2011fb} (not shown) in the same
centrality range. In more peripheral collisions, the STAR data in
Ref.~\cite{Sorensen:2011fb} tends to increase as seen with the AMPT
$v_3\{2\}$ results. This suggests that while $v_3\{2\}$ measurements
for $N_{\mathrm{part}}>100$ are dominated by the correlation of
particles with the participant plane, in more peripheral collisions
$v_3\{2\}$ begins to reflect correlations related to mini-jet
structure similar to that in $p+p$ collisions.

\begin{figure}[htb]
\centering
 \resizebox{0.49\textwidth}{!}{\includegraphics{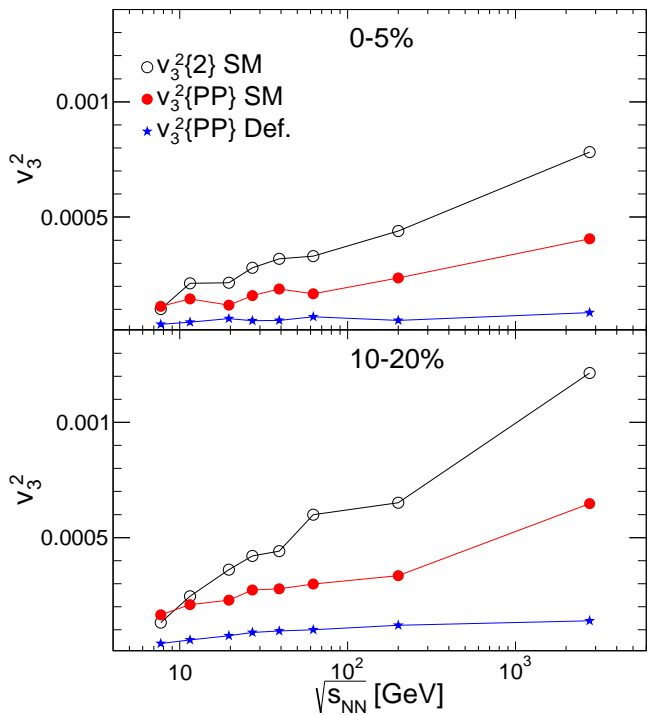}}
 \caption[]{ The $\sqrt{s_{_{NN}}}$ dependence of $v_3^2\{2\}$ (SM),
   $v_3^2\{\mathrm{PP}\}$ (SM), and $v_3^2\{\mathrm{PP}\}$ (Default)  for two
   different centrality intervals. }
\label{f5}
\end{figure}

In figure~\ref{f5} we show the AMPT results for the variation of
$v_3^2\{2\}$ and $v_3^2\{\mathrm{PP}\}$ with $\sqrt{s_{_{NN}}}$ from 7.7 GeV to
2.76 TeV for two centrality intervals. The results on $v_3^2\{\mathrm{PP}\}$
using the default setting for AMPT are very small and well below the
preliminary data presented by STAR. The $v_3^2\{\mathrm{PP}\}$ SM results
decrease rather smoothly with decreasing energy but still have an
appreciable value down to 7.7 GeV. The calculations for $v_3^2\{2\}$
SM have the same value as the $v_3^2\{\mathrm{PP}\}$ SM at 7.7 and 11.5
GeV. This again indicates that within this model, non-flow from
minijets has a negligible impact on two-particle correlations at the
lowest energies measured in the RHIC beam energy scan. Above those
energies, the difference between $v_3^2\{2\}$ and $v_3^2\{\mathrm{PP}\}$ grows
substantially. It will be interesting to see if the experimental data
on $v_3$ follows the same trend as AMPT SM all the way down to 7.7
GeV where non-flow from minijets can be neglected. It will be most
interesting to see if data eventually drops down to the values
predicted by the AMPT Default model. Estimates of the Bjorken energy
density~\cite{Bjorken:1982qr} compared to the Lattice QCD estimates for the critical energy
density~\cite{Petreczky:2004xs} suggest that this may not happen until below 7.7
GeV~\cite{phenixbj}. The calculations presented in this paper provide a
base-line with which to compare future experimental data.

\section{Viscosity Estimates}

In Ref.~\cite{Xu:2011fi}, the authors estimate the shear viscosity to
entropy ratio $\eta_s/s$ based on kinetic theory using
\begin{equation}
\eta_s/s \approx \frac{3\pi}{40\alpha_s^2} \frac{1}{\left(9+\frac{\mu^2}{T^2}\right)\ln\left(\frac{18+\mu^2/T^2}{\mu^2/T^2}\right)}
\label{visc}
\end{equation}
where $\mu$ is the screening mass of a gluon in the QGP and $\alpha_s$
is the QCD coupling constant. Both are input parameters for the AMPT
model. In Ref.~\cite{Xu:2011fe}, it is shown that parameter set B
($\mu=3.2$ fm$^{-1}$ and $\alpha_s=0.33$) provides a good description of
the $v_2$ and $v_3$ data at $\sqrt{s_{_{NN}}} =$ 200 GeV and 2.76
TeV. These parameters and equation~\ref{visc} then yield the
$\eta_s/s(T)$ as shown in figure~\ref{f7} with the label Set B. For
the estimated initial temperatures (378 MeV) at top RHIC energy,
$\eta_s/s$=0.38, far above the ADS/CFT or quantum lower bound of
$1/4\pi$~\cite{gyulassy,Kovtun:2004de} and also above most estimates
from hydrodynamic models.

\begin{figure}[htb]
\centering
\resizebox{0.5\textwidth}{!}{\includegraphics{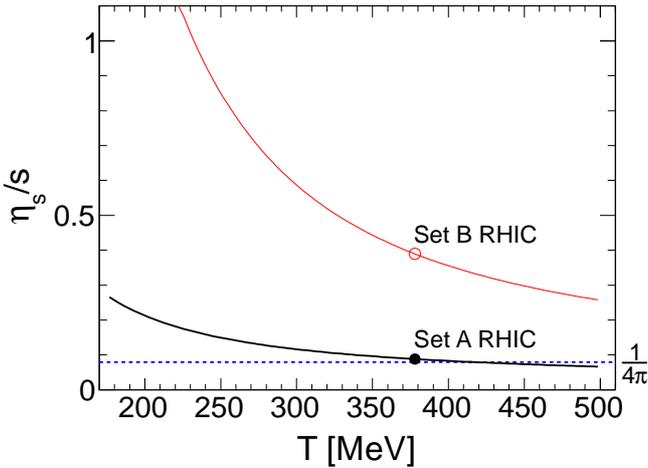}}
\caption[]{ The temperature dependence of $\eta_s/s$ according to
  eq.~\ref{visc} for two parameter sets.}
\label{f7}
\end{figure}

In this study, we have used parameter set A ($\mu=1.8$ fm$^{-1}$ and
$\alpha_s=0.47$) corresponding to a larger partonic scattering cross
section of 10 mb rather than 1.5 mb for set B. We find a good
description of the data with set A by turning of initial and final
state radiation in HIJING. This reduces the initial entropy production
and multiplicity but in a way that matches the multiplicity at all the
energies studied without varying any other parameters. The smaller
initial multiplicity compensates for the larger cross sections so that
the data are still well described. The $\eta_s/s(T)$ estimated from
this parameter set is labeled Set A in figure~\ref{f7}. For set A,
$\eta_s/s$ at 378 MeV is 0.088 which is very close to the ADS/CFT
conjectured lower bound. We conclude therefore that the AMPT model
can give a good description of the $v_2$ and $v_3$ data with a wide
range of $\eta_s/s$ values and that it's crucial to understand
the initial entropy production in order to extract the correct value
of $\eta_s/s$ in the QGP phase.

\section{Summary}
 
In this paper, we have presented AMPT SM and Default calculations for
$v_2$ and $v_3$. The primary purpose of these calculations is to
provide a reference for measurements of the beam energy dependence of
$v_2$ and $v_3$. We found that we can describe RHIC and LHC data on
multiplicity, $v_2$ and $v_3$ by turning off initial and final state
radiation in HIJING (reducing the initial entropy) but keeping
relatively large cross-sections in the QGP phase. 
Whereas a previous study found a good description
of data using a much smaller cross-section implying a much larger
value for $\eta_s/s$, our studies with a larger cross-section implies
a ratio of viscosity to entropy much closer to the ADS/CFT conjectured
lower bound. We have also studied how $v_2(p_T)$ changes from 7.7 GeV
to 2.76 TeV. We find that within this model, $v_2(p_T)$ changes very
little across the whole energy range studied, consistent with what is
observed in data. We also find that AMPT reproduces the experimental
observation that $v_3/\varepsilon_3 \propto
\sqrt{N_{\mathrm{part}}}$. These experimental observations therefore
seem to be understandable without major changes to our
description of heavy ion collisions and a subsequentx nearly perfect liquid QGP phase. Our studies of
the centrality and beam energy dependence of $v_2$ and $v_3$ with SM
and Default settings provide a comparitive base-line for studies of
$v_2$ and $v_3$ in the RHIC beam energy scan.

\textbf{Acknowledgements} We would like to thank Zi-Wei Lin, Jun Xu
and Che-Ming Ko for their guidance in carrying out the AMPT
studies. This work is supported by the U.S. Department of Energy under
contract DE-AC02-98-CH10886. Financial support from the Department 
of Science and Technology, Government of India, is gratefully acknowledged.

\end{document}